\documentclass{aastex}
\usepackage{spr-astr-addons}
\usepackage{url}
\usepackage{graphicx}

\RequirePackage{color}

\newcommand{\emaila}{torabi@ipm.ir}

\begin{document}

\title{A Simple Procedure to Extend the Gauss Method of Determining Orbital Parameters from Three to N Points}
\slugcomment{}

\author{Taghi Mirtorabi}
\affil{Department of Physics, Alzahra University, Tehran, IRAN \\ School of Astronomy, Institute for Research in Fundamental Sciences (IPM), Tehran, IRAN}
\email{\emaila}

\begin{abstract}
A simple procedure is developed to determine orbital elements of an object orbiting in a central force field which contribute more than three independent celestial positions. By manipulation of formal three point Gauss method of orbit determination, an initial set of heliocentric state vectors ${\bf r}_i$ and $\dot{\bf{r}}_i$ is calculated. Then using the fact that the object follows the path that keep the constants of motion unchanged, I derive conserved quantities by applying simple linear regression method on state vectors ${\bf r}_i$ and $\dot{\bf{r}}_i$. The best orbital plane is fixed by applying an iterative procedure which minimize the variation in magnitude of angular momentum of the orbit. Same procedure is used to fix shape and orientation of the orbit in the plane by minimizing variation  in total energy and  Laplace Runge Lenz vector. The method is tested using simulated data for a hypothetical planet rotating around the sun.
\end{abstract}


\section{Introduction}
\label{intro}
Orbit determination is a rather old problem, dating back to late eighteenth century when Laplace developed his solution \citep{Taff} and early nineteenth when Giuseppe Piazzi discovered Ceres and the famous young mathematician, Carl Friedrich Gauss, developed an efficient method of orbit determination \citep{Guas1} to recover the dwarf planet after its reappearance. Launching Sputnik in 1957 arose the need for orbit determination. \cite{Weiff} and \cite{Guier} used data from doppler tracking of Sputnik satellite which formed the basis of the transit system. By development in the performance of camera and radio doppler tracking systems in 1960s, the orbit precession improved to about 10-20 meter. Advances in laser technology in 1970s have raised the accuracy to few centimeter in altitude. Today orbit precessions are routinely in the 2 to 5 cm range \citep{Tapley1}.

In response to advances in observational accuracy, orbit determination methods also were improved. \cite{Kozai} developed a first order theory using Lagrange's planetary equations. This method was the basis of the Smithsonian Astrophysical Observatory Differential Orbit Improvement Program that was used to analyze very accurate Baker-Nunn camera observations and now used as a basis for NASA GEODYN program for precision geophysics application. \cite{Brouwer} adapted the Hill-Brown lunar theory to low-Earth satellite problem and developed a method which uses  mean orbital elements and include inclination and eccentricity as power series. \cite{Lydanne} modified Brouwer's  method to handle the singularities of eccentricity and inclination. Using osculating orbital elements, \cite{Kaula} developed a theory in Keplerian orbital element space. Kaula theory incorporated third body, resonance and tidal effects. It did not suffer from singularities and handled more general cases. The theory developed by Brouwer and modified and improved by Lydanne and Kaula is now the basic analytical theory used in astro-dynamical orbit determination codes.

The orbit determination method is now a mature topic which is subjects of many classical text books,\citep{Vallado,Danby,Tapley2,Taff,Montenbruck}. In the heart of the method there is a force model which incorporates variety of external interactions  including gravity, atmospheric drag, solar radiation pressure, third-body perturbations, Earth tidal effects, and general relativity. Using statistical procedures the method searches the parameter space for a reference orbit which closely resembles the observed one by minimizing a performance index proportional to sum of square of errors. This least squares criterion was first proposed by \cite{Gauss} and is commonly used today, \citep{Lawson,Bjorck}.  A short review of developments in orbit determination can be found in \cite{Vetter}.

In this paper a simple algorithm of orbit determination which implement more than three observations is presented. The method is based on linear relations in the definition of conserved quantities. By applying simple regression a constant of motion emerge itself as a constant of linearity which contribute all observations. Gronchi, Dimare, and Milani, also addressed the problem of orbit determination using two body integrals, \citep{milani}. They used energy and angular momentum to write a system of polynomial equations for the topo-centric distance and the radial velocity to calculate final solutions.

Recovering six orbital parameters needs at least six independent observations. In usual astronomical situations where there is no capability to measure state vector directly, three independent positioning is needed to accumulate the six orbital parameters. Modern observational instrumentation made possible to acquire hundreds of independent positions  for a celestial body in matter of hours. There  is no definite way to choose three among these huge data set to calculate the most precise orbital elements. Gauss method requires these three independent positions to be close to better approximate the law of areas, but if closeness is comparable to the observational error, the plane of the orbit would be ill-defined. Wide spaced positions would better illustrate the plane of the orbit but failed to be a good candidates to satisfy the Gauss approximation.

Here after in this paper I assume that an orbit in a two body central force problem can be fixed by five parameters which are two directional cosine of angular momentum vector represented by a unite vector $\bf{n}$ orthogonal to plane of the orbit, two geometric parameter which specify shape of the orbit in orbital plane which are semi major axis and eccentricity and one parameter which specify orientation of the orbit in the plane. To fix position of the body on the orbit we need one more parameter which is true or mean anomaly of the object in some specified instant of time. For the objects orbiting in the solar system I assume  earth sidereal year and astronomical unit as unit of time and distance respectively then $ \mu = G(m_1 + m_2) \simeq 4\pi^2$, where $m_1$ and $m_2$ are the sun and planet mass respectively $(m_1\gg m_2)$.

\begin{figure}[t]
  \includegraphics[width=8cm]{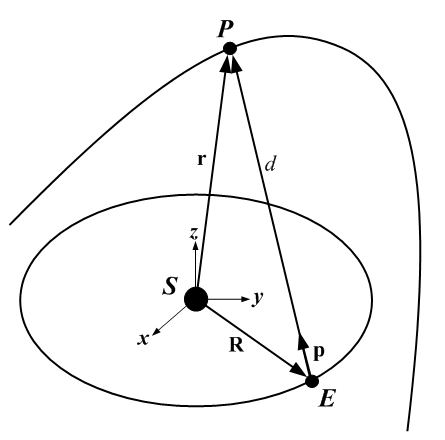}\\
  \caption{An observation from Earth ($E$) is correspond to a unit vector $\bf{p}$ to the object ($P$) with heliocentric state vector ${\bf r}$ and geocentric distance $d$.   }
  \label{fig1}
\end{figure}

\section{Plane of the orbit}

A major property of central forces is that they are torque less, keeping angular momentum conserved and produces two dimensional planar motion. I specify the plane of the motion with a unit vector ${\bf n}$ directed along the constant angular momentum vector. To determine ${\bf n}$, first I use Gauss method to estimate an initial value for ${\bf n}$ by dividing observations to groups of three. Then I try to improve ${\bf n}$ by searching for a plane which gives a constant angular momentum for all observations.

Suppose $N$ geocentric successive observations are made at times $t_1, t_2, t_3, ..., t_N$  where $$t_1 < t_2 < t_3 , ...,t_{N-1} < t_N.$$ Each observation is identified by a unit vector ${\bf p}_i$. As shown in Fig 1, the heliocentric state vector ${\bf r}_i$ of the object is related to observer position ${\bf R}_i$ and the observation ${\bf p}_i$ by
\begin{equation}\label{tri}
   {\bf r}_i = {\bf R}_i + d_i{\bf p}_i
\end{equation}
  where $d_i$ is geocentric distance of the object which in a telescope base observation is unknown. Here after by $\{X\}$ I denote a set of $N$ quantity $X_i$s each correspond to a single observation at $t_i$; for example $\{\bf{p}\}$ denote all geocentric unit vectors which represent the whole observation set. In absence of perturbation of other planets,  conservation of angular momentum implies that all state vectors $\{\bf{r}\}$, are lie in the same plane then
\begin{equation}\label{cop}
    {\bf r}_i = c_{i-1}{\bf r}_{i-1} + c_{i+1}{\bf r}_{i+1}
\end{equation}
where $c_{i-1}$ and $c_{i+1}$ are related to the area swept by state vectors ${\bf r}_{i-1}$ and ${\bf r}_i$. In the classical Gauss method, Kepler second law is employed to estimate $c_{i-1}$ and $c_{i+1}$. Assuming that the time between observations ${\bf p}_{i-1}$, ${\bf p}_{i}$, and ${\bf p}_{i+1}$ are small then $c_{i-1}$ and $c_{i+1}$ can be approximated by
\begin{equation}\label{gus}
    c_{i-1} \simeq \frac{t_{i+1} - t_i}{t_{i+1} - t_{i-1}} \ \ , \ \ c_{i+1} \simeq - \frac{t_{i-1} - t_i}{t_{i+1} - t_{i-1}}
 \end{equation}
 Substituting eq.(\ref{tri}) to eq.(\ref{gus}) will lead us to a set of linear equations for $d_i$s
 \begin{eqnarray}\label{di}
  & & c_{i-1}d_{i-1}{\bf p}_{i-1} - d_i{\bf p}_i + c_{i+1}d_{i+1}{\bf p}_{i+1} \nonumber \\
  & & =  -c_{i-1}{\bf R}_{i-1} + {\bf R}_i - c_{i+1}{\bf R}_{i+1}
 \end{eqnarray}
 which can be solved for $d_i$ by multiplying both side of eq.(\ref{di}) with $({\bf p}_{i-1} \times {\bf p}_{i+1})$, \cite{curtis}
 \begin{equation}\label{di2}
    d_i = \frac{[c_{i-1}{\bf R}_{i-1} - {\bf R}_{i} + c_{i+1}{\bf R}_{i+1}].({\bf p}_{i-1} \times {\bf p}_{i+1})}{{\bf p}_i.({\bf p}_{i-1} \times {\bf p}_{i+1})}.
 \end{equation}
 Putting back $\{d \}$ in eq.(\ref{tri}) will lead us to the set of all state vectors $\{\bf{r}\}$. The unit vector $\bf{n}$ can be determined by fitting a plane to the  $\{\bf{r}\}$ vector set. Least square plane fit method will give us two solution, $\bf{n}$ and $-\bf{n}$. The relevant $\bf{n}$  is the one which represent correct sense of the motion. A solution with $n_z > 0$ represents a prograde rotation and in a retrograde rotation, solution with  $n_z < 0$ should be chosen.

 To improve the unit vector ${\bf n}$  the state vectors $\{\bf{r}\}$ are regenerated to be all lie in this adopted plane by using the relation
 \begin{equation}\label{cross}
    {\bf r}_i = \frac{ {\bf n} \times {\bf A}_i}{({\bf n}.{\bf p}_i)}
 \end{equation}
 where
 \begin{equation}\label{A}
    {\bf A}_i =  {\bf R}_i \times {\bf p}_i.
 \end{equation}
 The velocity vectors $\{\dot{{\bf r}}\}$ can also be derived by differentiating of eq.(\ref{cross})
\begin{equation}
    \dot{{\bf r}}_i = \frac{ {\bf n} \times \dot{{\bf A}}_i}{({\bf n}.{\bf p}_i)} - \frac{({\bf n}.\dot{{\bf p}}_i)({\bf n} \times {\bf A}_i)}{({\bf n}.{\bf p}_i)^2}
\end{equation}
vector  product of ${\bf r}_i$ and $\dot{{\bf r}}_i$  lead us to the specific angular momentum of the orbit
\begin{equation}\label{ang}
    {\bf r}_i \times \dot{{\bf r}}_i = [\frac{{\bf n}.({\bf A}_i \times \dot{{\bf A}}_i)}{({\bf n}.{\bf p}_i)^2}] {\bf n} = L_i {\bf n}
\end{equation}
Equation (\ref{ang}) implies that the magnitude of the specific angular momentum $L_i$ for each single observation can be determined directly from the observation and the unit vector $\bf{n}$
\begin{equation}\label{ang2}
    L_i = \frac{{\bf n}.({\bf A}_i \times \dot{{\bf A}}_i)}{({\bf n}.{\bf p}_i)^2}
\end{equation}
$L_i$ is a conserved quantity which means, in absence of perturbation and observational errors, the best orbital plane ($\bf{n}$) for observations $\{\bf{p}\}$ is the one which keep $\{L\}$ constant. This lead us to a procedure to improve $\bf{n}$. The procedure is based on the fact that constant $\{L\}$ implies a linear relation between ${\bf n}.({\bf A}_i \times \dot{{\bf A}}_i)$ and $({\bf n}.{\bf p}_i)^2$. By implementing a simple linear regression method we can calculate two constants $L$ and $Q$ which minimize the sum of squared residuals such that
\begin{equation}\label{eqn}
    {\bf n}.({\bf A}_i \times \dot{{\bf A}}_i) = L ({\bf n}.{\bf p}_i)^2 + Q
\end{equation}
Here $L$ manifest itself as the best estimate of constant specific angular momentum contributing all observations. $Q$ can also be taken as an estimate of error in $\bf{n}$.  Setting $Q = 0$ we can rewrite equation (\ref{eqn}) as
\begin{equation}\label{it}
    {\bf n}.[({\bf A}_i \times \dot{{\bf A}}_i) -  L ({\bf n}.{\bf p}_i){\bf p}_i] = 0
\end{equation}
which indicates that in absence of errors or approximations the vector set
 \begin{equation}\label{g}
    \{ {\bf G} \} =\{ ( {\bf A} \times \dot{{\bf A}}) -  L (\bf{n}.\bf{p})\bf{p} \}
 \end{equation}
 are lie in the plane of the orbit. Then a better estimate of $\bf{n}$ can be obtained by fitting a plane to the $\{\bf{G}\}$ vector set and checking if the new $\bf{n}$ (the normal vector to the plane which represent correct sense of the motion) will reduce the variation in $\{L\}$. Equation (\ref{it}) can be manipulated iteratively to see if a statistical factor like standard deviation ($\sigma_L$) of $\{L\}$ reduces to a minimum. It is not obvious that this procedure always converge to a minimum for $\sigma_L$, but if does, the new $\bf{n}$ will give us a better solution for the plane of the orbit. It seems that reduction in $\sigma_L$ have a tendency to compensate all approximations we did before when we estimate $c_is$ in eq.(\ref{gus}) and in eq.(\ref{eqn}) when we calculate $\dot{\bf{A}}_i$ which normally might be an Euler differentiation.

\section{Shape of the orbit}

By the way, it is evident that conservation of total energy implies another linear relation which leads us to fix the shape of the orbit. In inverse square central force problem conservation of energy implies a linear relation between $\{\dot{\bf{r}}^2\}$  and $\{\frac{1}{r}\}$.
\begin{equation}\label{en}
    \dot{\bf{r}}^2_i = K \frac{1}{r_i} + E
 \end{equation}
 where $K$ and $E$ are constants of linearity and can be determined by fitting a line to both set of $\{\dot{\bf{r}}^2\}$  and $\{\frac{1}{r}\}$. $E$ is proportional to total energy of the orbit. For  ($E \neq 0$)  the orbit will be an ellipse or hyperbola  with semi major axis
\begin{equation}\label{orbit1}
     a = -\frac{K}{2E}.
\end{equation}
If $E$ vanishes then the orbit is a parabola ($e = 1$) with perihelion distance
\begin{equation}\label{orbit2}
    p = \frac{L^2}{K}.
\end{equation}

The last constant of the motion is Laplace Runge Lenz vector or eccentricity vector, $\bf{e}$ which lies in the plane of the orbit, directed to perihelion with  magnitude equal to the eccentricity. This vector can be expressed as
\begin{equation}\label{e}
    ({\bf r}_i \times \dot{{\bf r}_i}) \times \dot{{\bf r}_i} + 4\pi^2\frac{{\bf r}_i}{r_i} = - 4\pi^2 {\bf e}.
\end{equation}
Eq.(\ref{e}) is another linear relation between two vector sets $\{(\bf{r} \times \dot{\bf{r}}) \times \dot{\bf{r}} \}$ and $\{\frac{\bf{r}}{r} \}$. Here we have a vector relation which means simple linear regression must be applied to each of three component separately. The result is two constants $P$ and $\bf{M}$ where
 \begin{eqnarray}
   ({\bf r}_i \times \dot{{\bf r}_i}) \times \dot{{\bf r}}_i|_x &=& P \frac{{\bf r}_i}{r_i}|_x + M_x \nonumber \\
   ({\bf r}_i \times \dot{{\bf r}_i}) \times \dot{{\bf r}}_i|_y &=& P \frac{{\bf r}_i}{r_i}|_y + M_y \nonumber \\
   ({\bf r}_i \times \dot{{\bf r}_i}) \times \dot{{\bf r}}_i|_z &=& P \frac{{\bf r}_i}{r_i}|_z + M_z \label{e2}
 \end{eqnarray}
 Comparing eq.(\ref{e2}) with eq.(\ref{e}) shows the constant $P$ must approach to $4\pi^2$ but vector $\bf{M}$ can lead us to the best value for the eccentricity vector $\bf{e}$
 \begin{equation}\label{e3}
    {\bf e} = \frac{1}{P}{\bf M}.
 \end{equation}
 None of eccentricity vector components are independent from other constants of motion except one. It is conventional to express this last constant of motion by the argument of perihelion ($\omega$) which is defined as the angle between $\bf{e}$ and ascending node. Having both $\bf{n}$ and $\bf{e}$ in hand the argument of perihelion can be calculated as
 \begin{equation}\label{ln2}
    \omega = \cos^{-1}[\frac{{\bf e}.(\hat{{\bf k}} \times {\bf n})}{e}]
 \end{equation}
 where $\hat{{\bf k}}$ is the unit vector in the $z$ direction.

 The orbit is fixed by $\bf{n}$, $a$, $e$, and $\omega$. The final step is to find position of the object on the orbit. For each single observation the true anomaly is
 \begin{equation}\label{nu}
    \theta_i = \cos^{-1}(\frac{{\bf e}.{\bf r}_i}{er_i})
 \end{equation}
 I assumed that the reader is aware of trigonometrical method to recognize that the angle $\theta_i$ is belong to which quadrant.

\section{ The algorithm}\label{alg}

The step by step procedure to calculate orbital elements are outlined here in a simple algorithm:
\begin{enumerate}
  \item Given a set of $N$ observations $\{\bf{p}\}$ taken in consecutive times $t_1, t_2, t_3, ..., t_N$  such that $t_1 < t_2 < t_3 , ...,t_{N-1} < t_N$. Calculate  $\{d\}$ using eq.(\ref{di2}).
  \item Calculate the state vectors set $\{\bf{r}\}$, eq. (\ref{tri}).
  \item Find an initial value  for $\bf{n}$ by fitting a plane to the state vectors set $\{\bf{r}\}$. Check if $\bf{n}$ represent sense of the rotation correctly by checking $sgn({\bf r}_i \times {\bf r}_j \mid_z^{j>i}) = sgn(n_z)$. If not invert the $\bf{n}$.
  \item Calculate $\{\bf{n}.(\bf{A} \times \dot{\bf{A}})\}$ and $\{(\bf{n}.\bf{p})^2\}$. Perform a simple linear regression and obtain $L$ and $Q$. \label{a1}
  \item Use equation (\ref{ang2}) to calculate $\{L\}$. Calculate $\sigma_L$.\label{a2}
  \item Calculate $\{\bf{G}\}$, then fit a plane to them. Find unit vector normal to the plane which correctly represent sense of the motion. Set this unit vector as new $\bf{n}$, go back to step (\ref{a1}) and repeat until $\sigma_L$ reduce to a minimum value. This gives you the two directional cosine of plane of the orbit which is expressed by the final $\bf{n}$.\label{a3}

  \item Regenerate all $\{\bf{r}\}$ using eq.(\ref{cross}).
  \item Calculate $\{\dot{\bf{r}}^2\}$  and $\{\frac{1}{r}\}$, perform a simple linear regression between them and obtain $K$ and $E$. Calculate semi major axis $a$ using equation (\ref{orbit1}) or (\ref{orbit2}).
  \item Calculate $\{(\bf{r} \times \dot{\bf{r}}) \times \dot{\bf{r}}\}$ and $\{\frac{\bf{r}}{r}\}$, perform three successive simple linear regression for each component to find $P$ and $\bf{M}$. Find eccentricity vector using eq.(\ref{e3}).
  \item Calculate argument of perihelion $\omega$ using eq.(\ref{ln2}).
  \item Calculate true anomaly $\{\theta \}$ using eq.(\ref{nu}).

\end{enumerate}

\section{Test of the method}
To test the procedure presented in this paper, a set of 100 consecutive observations, $\{\bf{p}$;$t\}$, for a hypothetical planet rotating around the sun in a elliptical orbit were generated. The orbit was assumed to be an ellipse with semi major axis and eccentricity of $a = 5 AU$ ($T = 11.2$ years) and $e = 0.4$ respectively. The plane of the orbit was titled such that $i = 30^\circ \ ;\  \Omega = 30^\circ$  which implies, for a prograde motion the unit vector orthogonal to the orbital plane as $$ {\bf n} = (\frac{1}{4}, -\frac{\sqrt{3}}{4}, \frac{\sqrt{3}}{2}).$$
The argument of perihelion or direction of eccentricity vector with respect to line of nodes was set up to $ \omega = 45^\circ$. A uniform circular rotation in the plane of the ecliptic  was assumed for the observer. A uniform random error with amplitude of 0.1 arcsec was added to observations to simulate observational errors. The observations was arranged to be evenly spread in 100 successive nights (one observation per night).

Following  the algorithm given in section \ref{alg}, first estimate of $\bf{n}$ was determined to be $$ {\bf n} = ( 0.2477, \   -0.4288, \ 0.8688)$$ which implies an error of $ \Delta n = (\Delta n_x^2 + \Delta n_y^2 + \Delta n_z^2)^{\frac{1}{2}} = 0.0055 $. Using this unit vector the relative standard deviation of $\{L\}$ was calculated to be  $\frac{\sigma_L}{L} = 0.01$. After ten iterations through steps \ref{a1}-\ref{a3}, relative standard deviation was reduced to a minimum equal to  $\frac{\sigma_L}{L} =  0.000174$. This reduces the error in $\bf{n}$ to $ \Delta n = 3.59 \times 10^{-5}$.  The final $\bf{n}$ was calculated to be $$ {\bf n} = ( 0.24998, \   -0.43299, \  0.86604).$$

Proceeding the algorithm lead us to $$ a = 4.986 \ AU ; \ e = 0.4017; \ \omega=45.586^\circ. $$

\section{Conclusion}
The basic idea of the method presented in this paper is that an object in a central force will follow the orbit which keep constants of the motion unaltered. In absence of non conservative forces like atmospheric drag and perturbation of third body, I rebuild the conservation equations as linear relation between state vectors with unknown constant coefficients. Then try to estimate these unknown coefficients by statistically accumulating all observations. Finally the standard osculating orbital parameters are connected to these coefficients. By the way, it is obvious that this method is only applicable to the situations where the orbit is clean (no drag) and the conservation relation could be written  as eq.(\ref{eqn}), eq.(\ref{en}) and eq.(\ref{e2}), (no perturbation). In a very real situation where all perturbative effects are present this method can be used as a simple extension of three point method of Gauss to estimate an initial value for orbital parameters.

The basic advantage of this method is simplicity. Usual statistical  orbit determination methods are using sophisticated matrix manipulations with matrix dimension ($N$) equal to number of observations. The processing time in these codes are proportional to $N^2$, instead in simple linear regression the processing is proportional to $N$. In this respect the method is appropriate for robotic telescopes searching for Near Earths Objects (NEO). These telescopes must estimate the orbital parameters in short arcs where Gauss three point method fails to define the plane of the orbit. By accumulating more than three observations, the method presented here can  reduce statistical error and reproduce better estimate for orbital elements.

\acknowledgments
This work is financially supported by the Alzahra university office of research and technology.

\nocite{*}
\bibliographystyle{alpha}

\end{document}